\documentclass[aps,prl,preprint,showpacs,superscriptaddress,groupedaddress]{revtex4}
\usepackage{latexsym}
\usepackage{natbib}
\usepackage{amsmath}
\usepackage{amsfonts}
\usepackage{amssymb}
\usepackage[T1]{fontenc}
\usepackage[dvips]{color,graphicx}   
\usepackage{subfig}
\usepackage[percent]{overpic} 
\usepackage{float}  

\begin{document}

\title{Triple Point in Correlated Interdependent Networks}
\author{L. D. Valdez} \affiliation{Instituto de Investigaciones
  F\'isicas de Mar del Plata (IFIMAR)-Departamento de F\'isica,
  Facultad de Ciencias Exactas y Naturales, Universidad Nacional de
  Mar del Plata-CONICET, Funes 3350, (7600) Mar del Plata, Argentina.}
\author{P. A. Macri} \affiliation{Instituto de Investigaciones
  F\'isicas de Mar del Plata (IFIMAR)-Departamento de F\'isica,
  Facultad de Ciencias Exactas y Naturales, Universidad Nacional de
  Mar del Plata-CONICET, Funes 3350, (7600) Mar del Plata, Argentina.}
\author{H. E. Stanley}
\affiliation{Center for Polymer Studies, Boston University, Boston,
  Massachusetts 02215, USA}
\author{L. A. Braunstein} \affiliation{Instituto de Investigaciones
  F\'isicas de Mar del Plata (IFIMAR)-Departamento de F\'isica,
  Facultad de Ciencias Exactas y Naturales, Universidad Nacional de
  Mar del Plata-CONICET, Funes 3350, (7600) Mar del Plata, Argentina.}
\affiliation{Center for Polymer Studies, Boston University, Boston,
  Massachusetts 02215, USA}

\begin{abstract}

Many real-world networks depend on other networks, often in nontrivial
ways, to maintain their functionality. These interdependent ``networks
of networks'' are often extremely fragile. When a fraction $1-p$ of
nodes in one network randomly fails, the damage propagates to nodes in
networks that are interdependent and a dynamic failure cascade occurs
that affects the entire system. We present dynamic equations for
two interdependent networks that allow us to reproduce the failure
cascade for an arbitrary pattern of interdependency. We study the
``rich club'' effect found in many real interdependent network systems
in which the high-degree nodes are extremely interdependent, correlating
a fraction $\alpha$ of the higher degree nodes on each network. We find
a rich phase diagram in the plane $p-\alpha$, with a triple point
reminiscent of the triple point of liquids that separates a
nonfunctional phase from two functional phases.

\end{abstract}


\maketitle

Real-world infrastructures that provide essential services such as
energy supply, transportation, and communications~\cite{Rin_01} can be
understood as interdependent networks.  Although this interdependency
enhances the functionality of each network, it also increases the
vulnerability of the entire system to attack or random failure
\cite{Ves_01}. In these interdependent infrastructures, the disruption
of a small fraction of nodes in one network can generate a failure
cascade that disconnects the entire system.

Failure cascades in real-world interdependent systems, such as the 2003
electrical blackout in Italy caused by failures in the
telecommunications network \cite{Ros_01}, are physically explainable as
abrupt percolating transitions \cite{Bul_02,Bul_01,Bul_12}.  In
Ref.~\cite{Bul_02}, the authors study the simplest case of two networks
$A$ and $B$ of the same size $N$ with random interdependent nodes.
Within each network the nodes are randomly connected through
connectivity links, and pairs of nodes of different networks are
randomly connected via one-to-one bidirectional interdependent links,
enabling the failures to propagate through the links in either
direction. The random failure of a fraction $1-p$ of nodes in one
network produces a failure cascade in both networks. As a consequence,
the size of the giant component (GC) of each network, i.e., the
still-functioning network within each network, dynamically decreases
until the system reaches a steady state. Reference~\cite{Bul_02}
describes the existence of a critical threshold $p_c$, which is a
measure of the robustness of the entire network, below which the size of
the functioning network within each network abruptly collapses as a
first-order percolating transition and above which these functioning
networks are preserved.

In many real systems, however, this interdependency is not fully
random~\cite{Ron_02,Yan_01}. Instead, nodes of different networks
connect to form a ``rich club'' in which a portion of high-degree
nodes in one network depends on corresponding high-degree nodes in
other networks.  This occurs in trading and finance networks in which
a well-integrated country in the global trade market is also
well-integrated in the financial system. Another example of the
non-trivial patterns of interdependency can be found in
telecommunication networks in which important nodes often acquire a
battery backup system in order to decrease their dependence on the
electrical supply network. To understand the effect of these realistic
features on failure cascades, some studies have focused separately on
the correlation between the degrees of interdependent nodes
\cite{Bul_01,Ron_02} and the random or targeted autonomization
\cite{Ron_01,Hua_01,Di_01,Sch_01}.  In these studies, the original
theoretical formalism \cite{Bul_02} is reformulated to take into
account these features.

In this Rapid Communication, we present a simple, unified theoretical framework that
allows us to describe the dynamics of failure cascades in interdependent
networks for an arbitrary interdependency between networks. We
apply our framework to interdependent heterogeneous networks when a
fraction $\alpha$ of the higher degree nodes is interdependent, and a
fraction $1-\alpha$ is randomly dependent.  Here $\alpha$ is a parameter
that controls the level of correlation and allows us to explore its
effect on system robustness.

We consider for simplicity, but without loss of generality, two
networks $A$ and $B$ in which the degree distribution of the
connectivity links is given by $P[k_{A}]$ and $P[k_B]$, where $k_A$
and $k_B$ are the connectivity links of nodes in $A$ and $B$
respectively. We define $q_{A}[k_A,k_B]$ ($q_B[k_A,k_B]$) as the
fraction of nodes in network $A$ ($B$) that depends on network $B$
($A$). When $q_{i}[k_A,k_B]=1$ (with $i=A,B$) the system is one-to-one
and all the interdependent links are bidirectional, and for
$q_{i}[k_A,k_B]<1$ a node in network $A$ ($B$) with degree $k_A$
($k_B$) is independent of the other network with a probability
$1-q_{i}[k_A,k_B]$, i.e., the link cannot transmit the failure to that
node. After a random failure of $1-p$ nodes in network $A$ that
triggers the process, at each stage $n$ of the failure cascade that
goes from $A$ to $B$, a node is considered functional if it belongs to
the GC of its own network and the others become dysfunctional because
they lose support. As $f_{An}$ ($f_{Bn}$) is the probability that
transversing a link, a node of the giant connected component is
reached in network $A$ ($B$) at stage $n$~\cite{Bra_01,New_01,Cal_01},
a node on network $A$ with degree $k_{A}$ is functional if it can be
reached on its own network with a probability
$p(1-(1-pf_{An})^{k_A})$.  This node will not be affected by the
failure cascade (a) if it is independent of network $B$ with a
probability $1-q_{A}[k_A,k_B]$, or (b) if it depends on network $B$,
but its interdependent node in $B$ is connected to the GC at the
previous stage with a probability
$q_{A}[k_A,k_B]\bigl(1-(1-f_{Bn-1})^{k_B}\bigr)$. The relative size
$\Psi_{n}$ of the GC of network $A$ at stage $n$ is then given by
\begin{eqnarray}\label{eq1}
&&\Psi_{n}=p\Biggl(\sum_{k_A=k_{\rm min}}^{k_{\rm max}}
\sum_{k_B=k_{\rm min}}^{k_{\rm max}}P[k_A,k_B](1-q_{A}[k_A,k_B])(1-(1-pf_{An})^{k_A})+\nonumber\\&&\sum_{k_A=k_{\rm min}}^{k_{\rm max}}
\sum_{k_B=k_{\rm min}}^{k_{\rm max}}P[k_{A},k_B]q_{A}[k_A,k_B](1-(1-pf_{An})^{k_A})(1-(1-f_{Bn-1})^{k_B})\Biggr), 
\end{eqnarray}
where $P[k_A,k_B]$ is the joint degree distribution for the
interdependent links. The first term in Eq.~(\ref{eq1}) takes into
account the functional nodes in $A$ with degree $k_{A}$ which do not
depend on network $B$ and the second term corresponds to the case
where functional nodes in network $A$ with degree $k_A$, depend on
functional nodes of network $B$ with degree $k_{B}$ at step
$n-1$. Here $f_{An}$ fulfills the self consistent equation
\begin{eqnarray}\label{eq2}
&&f_{An}=\sum_{k_A=k_{\rm min}}^{k_{\rm max}}\sum_{k_B=k_{\rm min}}^{k_{\rm max}}
\frac{k_AP[k_A,k_B]}{\langle k_A \rangle}\left(1-q_{A}[k_A,k_B]\right)
(1-(1-pf_{An})^{k_A-1})+\nonumber\\&&\sum_{k_A=k_{\rm min}}^{k_{\rm max}}\sum_{k_B=k_{\rm min}}^{k_{\rm max}}
\frac{k_AP[k_{A},k_B]}{\langle k_A \rangle}q_{A}[k_A,k_B]
(1-(1-pf_{An})^{k_A-1})(1-(1-f_{Bn-1})^{k_B}).
\end{eqnarray}
Similarly, at stage $n$ the relative size $\phi_{n}$ of the GC of
network $B$ is given by
\begin{eqnarray}\label{eq3}
&&\phi_{n}=\sum_{k_A=k_{\rm min}}^{k_{\rm max}}\sum_{k_B=k_{\rm min}}^{k_{\rm max}}P[k_A,k_B](1-q_{B}[k_A,k_B])(1-(1-f_{Bn})^{k_B})+\nonumber\\&&p\sum_{k_A=k_{\rm min}}^{k_{\rm max}}\sum_{k_B=k_{\rm min}}^{k_{\rm max}}P[k_{A},k_B]q_{B}[k_A,k_B](1-(1-pf_{An})^{k_A})(1-(1-f_{Bn})^{k_B}),
\end{eqnarray}
where $f_{Bn}$ satisfies the self-consistent equation
\begin{eqnarray}\label{eq4}
&&f_{Bn}=\sum_{k_A=k_{\rm min}}^{k_{\rm max}}\sum_{k_B=k_{\rm min}}^{k_{\rm max}}\frac{k_BP[k_A,k_B]}{\langle k_B \rangle}(1-q_{B}[k_A,k_B])(1-(1-f_{Bn})^{k_B-1})+\nonumber\\&&p\sum_{k_A=k_{\rm min}}^{k_{\rm max}}\sum_{k_B=k_{\rm min}}^{k_{\rm max}}\frac{k_BP[k_{A},k_B]}{\langle k_B \rangle}q_{B}[k_A,k_B](1-(1-pf_{An})^{k_A})(1-(1-f_{Bn})^{k_B-1}).
\end{eqnarray}

Note that in the r.h.s of Eq.~(\ref{eq4}) $f_{Bn}$ is not multiplied
by $p$, since we assume that the initial failure of $1-p$ nodes occurs
only in network $A$.

In the steady state, i.e., for $n\to \infty$, $\Psi_{n}\approx
\Psi_{n-1}$ and $\phi_{n}\approx \phi_{n-1}$, thus $\Psi_{n}$ and
$\phi_{n}$ converge to $\Psi_{\infty}$ and $\phi_{\infty}$,
respectively. Our equations for the steady state were obtained by Son
$et al.$~\cite{Son_01} for uncorrelated interdependent networks and used
by Baxer $et al.$~\cite{Bax_01} to explain the origin of the avalanche
collapse.

We introduce here a correlated interdependency model, in which
interdependent links are connected bidirectionally and one-to-one
($q_{A}[k_A,k_B]=q_{B}[k_A,k_B]=1$), and a fraction $\alpha$ of the
higher degree nodes are fully correlated. This extends the ``rich
club'' concept~\cite{Col_01,Xu_01} to interdependent
networks. Assuming that the degree distribution of both networks is
the same, the joint degree distribution $P[k_A,k_B]$ is given by:

\begin{eqnarray}\label{cpattern}
P[k_A,k_B]=\left\{
\begin{array}{ll}
P[k_A]P[k_B]/(1-\alpha), &  k_A < k_S,k_B < k_S,\\
(1-w)P[k_S]P[k_B]/(1-\alpha),  & k_A=k_S,\; k_B< k_S,\\
(1-w)P[k_A]P[k_S]/(1-\alpha),  & k_B=k_S,\;k_A < k_S,\\
(1-w)^{2}P[k_S]P[k_S]/(1-\alpha)+ w\;P[k_S],  & k_A=k_B=k_S,\\
P[k_A]\delta_{k_A,k_B},  & k_{S} < k_A,\;k_{S} < k_B.
\end{array}
\right.
\end{eqnarray}
Here $k_S$ is the degree above which a fraction $\alpha$ of
interdependent nodes are correlated, and $w$ is the fraction of
correlated nodes with degree $k_S$ such that $wP[k_s]+\sum_{k=k_{S}+1}^{k_{\rm max}}P[k]=\alpha$

\begin{figure}[H]
\begin{center}
\vspace{0.5cm}
 \includegraphics[scale=0.30]{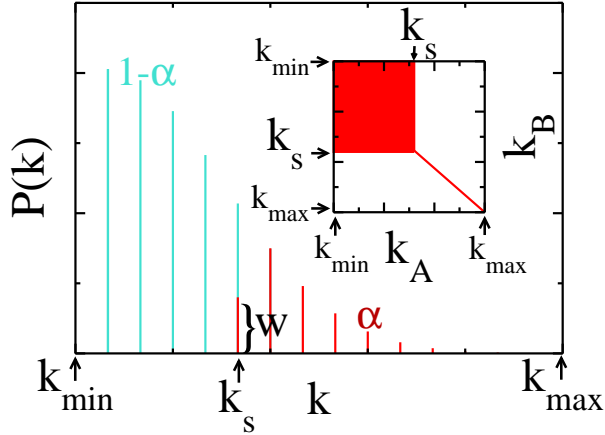}
  \vspace{0.5cm}
\end{center}
\caption{(Color online) Schematic of the degree distribution used to
  correlate the interdependent networks. If $k_s$ is the minimal
  degree for which the nodes are correlated, $\alpha$ represents the
  fraction of correlated interdependent nodes denoted by red, and
  $w$ is the fraction of interdependent correlated nodes with degree
  $k_s$. The light blue region represents the fraction $1-\alpha$ of
  uncorrelated nodes. In the inset we show with red color the
  pairs of interdependent nodes with degree $k_{A}-k_{B}$ present in
  this model.}
\label{fig.esq}
\end{figure}

In Eq.~(\ref{cpattern}), the factor $1-\alpha$ takes into account that a
fraction of nodes in two different networks with degree at and below
$k_S$ are randomly connected. In Fig.~\ref{fig.esq} we show
schematically the model used to correlate the degrees between
interdependent nodes and in the inset we show the pairs of
interdependent nodes with degree $k_{A}-k_{B}$.

As $P[k_A,k_B]=P[k_B,k_A]$ and by the symmetry of
Eqs.~(\ref{eq2}) and~(\ref{eq4}) in the steady state ($n\to \infty$),
$pf_{A\infty}=f_{B\infty}\equiv f_{\infty}$, and the self-consistent
equations reduce to
\begin{eqnarray}\label{eq5}
&&f_{\infty}=p\sum_{k_A=k_{\rm min}}^{k_{\rm max}}\sum_{k_B=k_{\rm min}}^{k_{\rm max}}\frac{k_BP[k_{A},k_B]}{\langle k \rangle}(1-(1-f_{\infty})^{k_A})(1-(1-f_{\infty})^{k_B-1}).
\end{eqnarray}

We apply this model to pure scale-free (SF) networks with $\lambda=2.5$,
$k_{\rm min}=2$ and maximal degree cutoff $k_{\rm max}=N^{1/2}$, with
$N=10^6$~\cite{Bog_01}. Here the finite cutoff mimics real networks in
which resources and energy are limited and nodes cannot have an
unbounded number of links~\cite{Ama_01}. 

\begin{figure}[H]
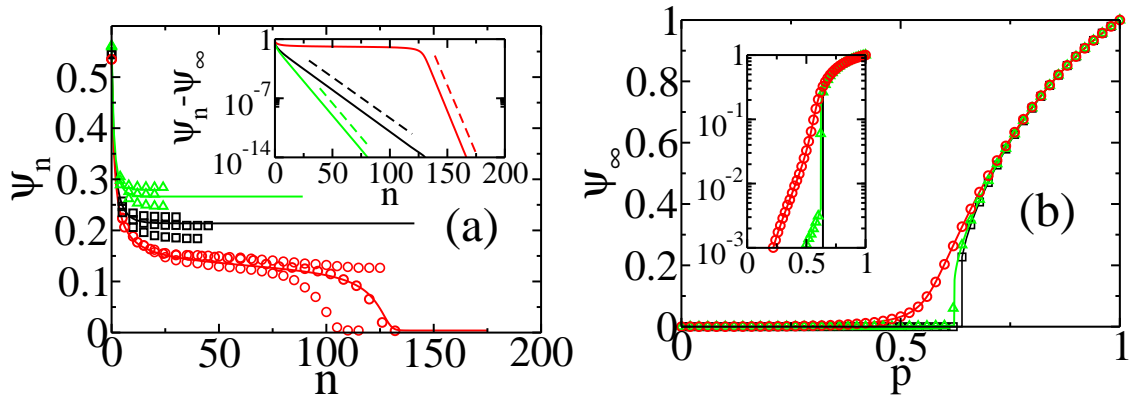

\begin{center}
\vspace{1cm}
  \begin{overpic}[scale=0.27]{temp.eps}
    \put(80,30){}
  \end{overpic}\vspace{0.5cm}
  \begin{overpic}[scale=0.27]{corr.eps}
    \put(80,30){}
  \end{overpic}\hspace{0.25cm}
\end{center}
\caption{(Color online) Cascade of failure on network $A$ for different values of
  $\alpha$ and $q=1$ on SF networks with $\lambda=2.5$ and $2\leq k \leq
  1000$. Figure a: $\Psi_{n}$ for $\alpha=0.01\%$ and $p=0.640$ (green,
  $\triangle$), $p=0.630$ (black, $\square$) and $p=0.622$ (red,
  $\bigcirc$) obtained from three single realizations of the simulations
  (symbols) and from Eqs.~(\ref{eq1})-(\ref{eq4}) (solid line). In the
  inset we show a log-linear figure of the exponential decay of
  $\Psi_{n}$ to $\Psi_{\infty}$. The dashed lines correspond to the
  exponential fit of the theoretical results with a characteristic
  time $\tau=2.70$, $\tau=4.5$ and $\tau=1.35$ for $p=0.640$, $p=0.630$
  and $p=0.622$, from top to bottom. Figure b: $\Psi_{\infty}$ as a
  function of $p$ obtained from simulations (symbols) and from
  Eqs.~(\ref{eq1})-(\ref{eq4}) (solid lines) for $\alpha=0.001\%$
  (black, $\square$), $\alpha=0.01\%$ (green, $\triangle$),
  $\alpha=0.1\%$(red, $\bigcirc$). In the inset we plot the main figure
  in log-linear scale in order to capture the abrupt collapse of the GC
  as explained in the text. The symbols are the average over 100 network
  realizations.}\label{fig.r_1}
\end{figure}

In Fig.~\ref{fig.r_1} we show the solution of the theoretical
equations (\ref{eq1})--(\ref{eq4}) and the simulation results for the
size of the GC of network A, $\Psi_{n}$, as a function of the stage
number $n$ (Fig.~\ref{fig.r_1}a) and $\Psi_{\infty}$ as a function of
the $p$ for different values of $\alpha$ (Fig.~\ref{fig.r_1}b)~\footnote{We assume that the system reaches
  a steady state when $\Psi_{n}-\Psi_{n+1}<10^{-18}$.}.

The figures show an excellent agreement between the theoretical
results and the simulations. In the temporal evolution,
Fig.~\ref{fig.r_1}a shows that a small variation in $p$ ($\Delta p
\approx 0.02$) can dramatically change the final size of the GC. The
inset of Fig.~\ref{fig.r_1}a shows that the approach of $\Psi_{n}$ to
$\Psi_{\infty}$ is exponential.  This behavior is due to the fact that
the number of iterations of $f_{n}$ in Eqs.~(\ref{eq2}) and
(\ref{eq4}) needed to reach the steady state is the same as the number
of iterations needed to find the fixed point of Eq.~(\ref{eq5}), in
which the approach of $\Psi_n$ to fixed point $\Psi_{\infty}$ is
exponential~\cite{Bul_02} and, as a consequence, the temporal
percolating dilution slows down.  
We can also see that at $p\approx
0.63$ the dilution rate decreases more quickly than for other values
of $p$, i.e., the size of the functional networks decays slowly,
indicating that there is time to intervene and prevent the collapse of
the GC.  This slow behavior around critical points are shown as peaks
in the number of iteration (NOI) steps needed to reach the steady
state, as we will show below.
Figure~\ref{fig.r_1}b shows that, as $\alpha$ increases, the system is
still functional for high initial failure values.  The critical
threshold $p_c$ at which the system is completly destroyed
decreases and thus the networks are more robust.  
\begin{figure}[H]
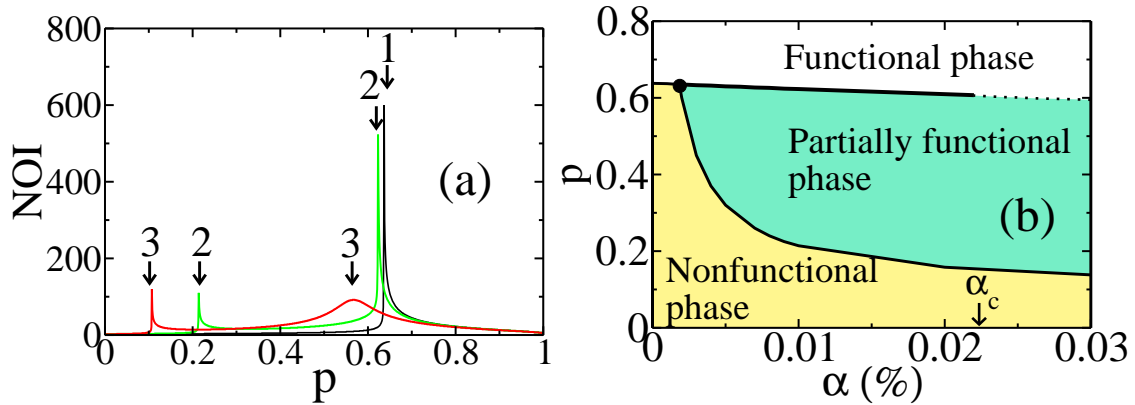

\centering
\vspace{0.5cm}
  \begin{overpic}[scale=0.27]{noi.eps}
    \put(80,30){}
  \end{overpic}\vspace{0.5cm}
  \begin{overpic}[scale=0.27]{phase.eps}
    \put(80,30){}
  \end{overpic}
\caption{(Color online) Figure (a): the NOI as a function of $p$,
  obtained from the iterations of Eqs.~(\ref{eq1})-(\ref{eq4}). The
  network parameters and the color are the same than in
  Fig.~\ref{fig.r_1}. The labels denote the position of the peaks for
  $\alpha=0.001\%$ (label $1$), $\alpha=0.01\%$ (label $2$),
  $\alpha=0.1\%$(label $3$). Figure (b): phase diagram in the plane
  $\alpha-p$: i) the light yellow area corresponds to the nonfunctional
  phase, $i.e$, $\Psi_{\infty}=\phi_{\infty}=0$, ii) the green area
  corresponds to a partial functional phase in which the size of the GC
  of both networks is $\lesssim 10^{-3}$ and iii) the white area
  corresponds to a functional phase where
  $\Psi_{\infty}=\phi_{\infty}\gtrsim 10^{-2}$. The black point on the
  left corresponds to the triple point. The solid lines represent the
  abrupt change on the network's sizes and the dotted line, which it is
  defined for $\alpha>\alpha_{c}$, represents a fast and continuous
  variation of $\Psi_{\infty}$ at $p_{c}^{+}$. }\label{fig.r_2}
\end{figure}
Note that, because we are using a finite degree cutoff when $\alpha
\to 1$, the threshold does not go to zero, but when $k_{\rm max}\to
\infty$ in SF networks with $\lambda \le 3$ and $\alpha=1$, $p_c\to 0$
in this limit~\cite{Bul_01}.

In order to demonstrate how correlation improves the robustness of the
networks in Fig.~\ref{fig.r_2}a we show the NOI of these systems. For
very low values of $\alpha$ there is only one peak at the critical
threshold $p_c$ that is related to a first order percolating
transition. Surprisingly, for increasing $\alpha$ (see the case of
$\alpha=0.01\%$ in the figure) there is another peak around the
threshold $p_c^{+}>p_{c}\equiv p_{c}^{-}$ at which the sizes of the
GCs decrease abruptly but, because the hubs support each other, the
functional networks are not destroyed, and the robustness of the
system against failure cascades is enhanced. For higher values of
$\alpha$ we also find that there is a sharp peak that corresponds to a
first order phase transition at $p=p_{c}^{-}$ and a rounded peak at
$p=p_{c}^{+}$ around which the size of the GC decreases continuously
with an increasing value of its derivative with respect to $p$,
$d\Psi_{\infty}/dp$ close to $p_{c}^{+}$. These findings suggest that
finite correlations generate a crossover between an abrupt and a
continuous-sharply decreasing in the sizes of the GCs.

Figure~\ref{fig.r_2}b shows the rich phase diagram in the $p-\alpha$
plane. Note that as $\alpha$ increases, the line of the first order
transition that separates a funtional GC phase from a nonfunctional
phase forks into two branches, generating a new phase characterized by a
small GC ($\lesssim 10^{-3}$).  Around this point small fluctuations in
the temporal evolution---or in the steady state---can induce an abrupt
change in the size of the GC, which is reminiscent of the instability of
the triple point of liquids where three phases
coexist~\cite{Sta_01}. The lower branch that emerges from the triple
point corresponds to the first order transition that separates
functional from nonfunctional phases. The upper one corresponds to the
second threshold where the dynamics slows down and, at
$\alpha=\alpha_{c}=0.0218\%$~\cite{Exxon_01}, the transition changes
from an abrupt variation to a rapid but continuous variation of
$\Psi_{\infty}(p)$. The small value of $\alpha_c$ indicates that a small
correlation of the highest degree nodes can avoid the abrupt change in
the size of the GC. We found the
same qualitative behavior for other SF networks with $2 < \lambda\leq
3$~\footnote{Close to $\lambda=2$ the robustness of the system is
  dominated by the divergence of the average degree, and the
  correlations have little effect on robustness of the systems.},
indicating that the triple point is characteristic of nontrivial
patterns of interdependency.

In summary, we have used a general framework to describe the
temporal behavior of failure cascades with any pattern of
interdependency links, and we have found a rich phase diagram for
degree-degree correlated interdependency with a triple point at which a
first order transition line splits into two first order lines with an
abrupt collapse of the sizes of the functional networks.  The agreement
between theory and simulations is excellent. Our framework can be
extended to study the dynamics of failure cascades and the robusteness
of networks with degree-degree correlation in their connectivity links
and in their multiple interdependent links, where we expect to find a
rich phase diagram~\cite{Val_01}. 

\section{acknowledgments}
 L.D.V, P.A.M and L.A.B thank UNMdP and FONCyT (Pict 0293/2008) for
 financial support. H.E.S thanks ONR (Grant N00014-09-1-0380, Grant
 N00014-12-1-0548), DTRA (Grant HDTRA-1-10-1- 0014, Grant
 HDTRA-1-09-1-0035), and NSF (Grant CMMI 1125290).
\bibliography{bib}

\begin{thebibliography}{24}
\bibitem[{Exx()}]{Exxon_01}
\bibinfo{note}{In order to calculate the critical point, we solve geometrically
  the Eq.~(\ref{eq5}) using the intersection between the identity function and
  the r.h.s of Eq.~(\ref{eq5}). A critical point $p_c$ corresponds to the value
  of $p$ at which the angle in the intesection point between the identity
  function and the r.h.s of Eq.~(\ref{eq5}) is minimun .}
\end{thebibliography}

\end{document}